\documentclass[11pt]{article}

\usepackage{euscript}
\usepackage{amssymb}
\usepackage{amsfonts}
\usepackage{amsbsy}
\usepackage{amsmath}
\usepackage{epsfig}
\usepackage{amsthm}
\usepackage{amscd}
\usepackage{amstext}

\usepackage[pdftex]{hyperref}

\def\ben{\begin{equation}}
\def\een{\end{equation}}
\def\half{{\textstyle{1\over2}}}

   \let\k=\kappa
  \let\n=\nu

\let\w=\omega    \let\L=\Lambda

\let\pa=\partial
\def\be{\begin{equation}}
\def\ee{\end{equation}}
\def\ba{\begin{array}}
\def\ea{\end{array}}

\def\dalemb#1#2{{\vbox{\hrule height .#2pt
       \hbox{\vrule width.#2pt height#1pt \kern#1pt
               \vrule width.#2pt}
       \hrule height.#2pt}}}

\newcommand{\bea}{\begin{eqnarray}}
\newcommand{\eea}{\end{eqnarray}}
\newcommand{\tr}{{\rm tr} }
\newcommand{\Tr}{{\rm Tr} }

\def\Z{{{\Bbb Z}}}

\def\IZ{\Z}

\begin{document}

\begin{flushright}
NSF-KITP-09-145\\
\end{flushright}

\begin{center}

\vspace{1cm} { \LARGE {\bf Black hole determinants and quasinormal modes}}

\vspace{1cm}

Frederik Denef$^{\sharp,  \natural, \flat}$, Sean A. Hartnoll$^{\sharp, \natural}$ and Subir Sachdev$^{\sharp, \natural}$

\vspace{0.8cm}

{\it ${}^\sharp$ Department of Physics, Harvard University,
\\
Cambridge, MA 02138, USA \\}
\vspace{0.5cm}

{\it ${}^\natural$ Kavli Institute for Theoretical Physics, University of California, \\
Santa Barbara, CA 93106, USA \\}
\vspace{0.5cm}

{\it ${}^\flat $ Instituut voor Theoretische Fysica, U Leuven, \\
Celestijnenlaan 200D, B-3001 Leuven, Belgium \\}

\vspace{0.6cm}

{\tt  denef, hartnoll, sachdev @physics.harvard.edu} \\

\vspace{2cm}

\end{center}

\begin{abstract}

We derive an expression for functional determinants in thermal spacetimes as a product over the corresponding quasinormal modes. As simple applications we give efficient computations of scalar determinants in thermal AdS, BTZ black hole and de Sitter spacetimes. We emphasize the conceptual utility of our formula for discussing `$1/N$' corrections to strongly coupled field theories via the holographic correspondence.

\end{abstract}
\pagebreak
\setcounter{page}{1}

\section{Determinants and quasinormal modes}

In a semiclassical quantisation of gravity the partition function can be written schematically as
\be\label{eq:generalZ}
Z = \sum_{g_\star} \det\left(-\nabla^2_{g_\star} \right)^{\pm1} e^{-S_E[g_\star]} \,.
\ee
Here $g_\star$ are saddle points of the Euclidean gravitational action $S_E$. We use $g_\star$ to collectively denote the metric and any other nonzero fields. The $\det\left(-\nabla^2_{g_\star} \right)$ term schematically denotes the product of determinants of all the operators controlling fluctuations about the background solution $g_\star$. As usual, the determinants appear on the numerator or denominator depending on whether the fluctuations are fermionic or bosonic respectively. This formula is readily generalised to correlators of operators by including appropriate sources in the gravitational action.

Originally developed with a view to elucidating quantum gravitational effects in our universe \cite{gibbonshawking}, the semiclassical approach to quantum gravity has gained a new lease of life through the holographic correspondence \cite{Maldacena:1997re, Witten:1998zw}. This is because the semiclassical limit in gravity corresponds to the large $N$ limit of a dual gauge theory. Almost all works have restricted attention to the leading order large $N$ result, the exponent in (\ref{eq:generalZ}). Exceptions to this last statement include \cite{Gubser:2002zh, Gubser:2002vv, Hartman:2006dy,diaz} who studied an effect due to the determinant term in pure Anti-de Sitter space. In our recent paper \cite{us} we have stressed the fact that the semiclassical determinant term in (\ref{eq:generalZ}), a $1/N^\#$ correction, will be important in the framework of `applied holography'. This is the case because there are interesting effects, in for instance thermodynamics and transport, which are either entirely absent at leading order or else swamped by physical processes that are not those of prime relevance.

In general it is technically challenging to compute determinants in nontrivial background spacetimes. Even when the Euclidean saddles are homogeneous the calculation can be involved. For example,
\cite{Volkov:2000ih} computed the gravitational determinant on $S^2 \times S^2$ to obtain the semiclassical rate of nucleation of black holes in de Sitter spacetime. For most cases of interest, often black hole spacetimes, the eigenvalues of the fluctuation operator cannot be determined explicitly and an evaluation of the determinant is difficult without resort to WKB or other approximations (although see for instance \cite{Howard:1984qp} and citations thereof).

In this paper we present and derive a formula for determinants in black hole backgrounds, indeed finite temperature spacetimes more generally, in terms of the quasinormal modes of the spacetime. For (complex) bosonic fields the expression appears in section \ref{sec:euclthermal}, equation (\ref{QNMfinal}), and takes the form 
\be\label{eq:statement}
\frac{1}{\det\left(-\nabla^2_{g_\star} \right)} = e^{{\rm Pol}} \prod_{z_\star} \frac{|z_\star|}{4 \pi^2 T} \left| \Gamma \left( \frac{i z_\star}{2 \pi T} \right)\right|^2 \,,
\ee
where $z_\star$ are the quasinormal frequencies, $T$ is the temperature and $e^{\text{Pol}}$ is a simple polynomial contribution, fixed by a finite number of local terms discussed in section \ref{sec:locex}. Equation (\ref{QNMfinalprodform}) gives an alternative formulation and (\ref{fermifinal}) the fermionic analog.

The quasinormal frequencies of a spacetime determine the time evolution of perturbations of a particular field about the background. They are the poles of the retarded Green's function for the field in question (see e.g. \cite{Ching:1994bd, Ching:1995tj}). In the holographic correspondence they are furthermore the poles in the retarded Green's function of the operator dual to the bulk field \cite{Son:2002sd}. While the pole with smallest imaginary part determines the thermalisation timescale of the field theory \cite{Horowitz:1999jd} (within a linear reponse regime, c.f. \cite{Chesler:2008hg}), more generally the poles of the retarded Green's function are basic physical quantites in the dual field theory. They could be considered the closest notion to well-defined excitations in the strongly coupled theory. Furthermore, over the years a vast effort has been invested in the computation of quasinormal frequencies in many black hole spacetimes, for a recent review see \cite{Berti:2009kk}. Much of the insight gained from these studies can now be translated into the physics of one loop determinants. In short, we expect formulae like (\ref{eq:statement}) to have the following uses:
\begin{itemize}

\item It allows exact computation of the determinant in certain cases.

\item If a single quasinormal mode contains the physics of interest we can isolate its contribution to the determinant.

\item Approximate (for instance WKB) expressions for the quasinormal frequencies give a method for approximating the determinant.

\item The quasinormal modes are natural physical quantities in the dual strongly coupled field theory which does not have weakly interacting quasiparticle excitations. Expressing the dynamics of the field theory in terms of these modes may lead to conceptual insights into strongly coupled theories.

\end{itemize}

Below we will give a couple of examples of the first of these uses, by (re)computing the one loop partition function of scalar fields in thermal AdS, BTZ and de Sitter backgrounds. In our paper \cite{us} we used the second fact in the above list; the non-analytic behaviour of a specific quasinormal frequency (in the $T \to 0$ limit) as a function of a magnetic field led to quantum oscillations.

Our objective in this paper is to give a leisurely exposition of (\ref{eq:statement}), together with simple examples that illustrate the power of analyticity arguments combined with quasinormal mode sums.
Although our main interest is in black hole spacetimes, we also consider thermal AdS and de Sitter space as simple examples that illustrate the approach. We briefly discuss the numerical evaluation of formulae like (\ref{eq:statement}) in section \ref{sec:wkb}. In section \ref{sec:generalize} we note that (\ref{eq:statement}) can be generalised to conserved quantum numbers other than the frequency, for instance the angular momentum quantum number $\ell$. In the discussion we mention various possible generalisations and physical applications.

\section{Warmup examples}

Before deriving a formula for general black hole spacetimes, we will consider some simple one dimensional examples to illustrate the basic idea. 

\subsection{Harmonic oscillator}

The simplest case is a harmonic oscillator.
The Euclidean path integral at finite temperature $T$ for a complex scalar in one dimension --- or in other words for the harmonic oscillator with two real degrees of freedom --- is
\begin{equation}
 Z(\kappa) = \int {\mathcal{D}}\phi \, {\mathcal{D}}\bar{\phi} \, e^{- \int_0^{1/T} d\tau \, \left( |{\pa_\tau \phi}|^2 + \kappa |\phi|^2 \right)} \, ,   
\end{equation}
where periodic boundary conditions are understood: $\phi(\tau) = \phi(\tau + 1/T)$. This reduces to the evaluation of a functional determinant:
\begin{equation}
 Z(\kappa) = N \left( \det \left(-\partial_\tau^2 + \kappa\right) \right)^{-1}  
 = N \prod_n (\w_n^2+\kappa)^{-1} \, .
\end{equation}
where $N$ is a normalization factor, the product is a product over eigenvalues, and 
we introduced the bosonic thermal frequencies
\be\label{eq:bosonic}
\w_n = 2 \pi n T \,, \qquad \text{(bosonic)} \,.
\ee
Absorbing an infinite $\kappa$-independent factor $\prod_n \w_n^{-2}$ in $N$ and using a well known infinite product formula gives
\begin{equation}
 Z(\kappa) = N' \, \frac{1}{\sinh^2 \frac{{\sqrt{\kappa}}}{2T}} \, .
\end{equation}
To fix $N'$ it is sufficient to note that in the small $T$ or large $\kappa$ limit, $Z(\kappa)$ only gets a contribution from the ground state energy $E_0$: $Z(\kappa) \simeq e^{-E_0/T}$. This correctly identifies $E_0 = \sqrt{\kappa}$ and implies $N'=1/4$, so finally
\begin{equation} \label{harmoscZfin}
 Z(\kappa) = \frac{1}{4\sinh^2 \frac{\sqrt{\kappa}}{2T}} = \sum_{n_1,n_2} e^{-\sqrt{\kappa} \left( 1 + n_1 + n_2 \right)/T} \, ,
\end{equation}
which is indeed the 2d harmonic oscillator partition function $Z = \Tr \, e^{-H/T}$.
 
Let us now evaluate the same determinant in a slightly more convoluted way, which however will readily generalize to the systems of interest to us.\footnote{This is essentially an adaptation of an argument in Coleman's lecture on instantons 
\cite{coleman}. For another application of similar ideas to the problem of computing determinants in AdS/CFT, see \cite{Hartman:2006dy}. For the general mathematical framework to which (a more rigorous version of) this type of reasoning belongs, see for example \cite{Young}.}

Consider the function $Z(\kappa)$ analytically continued to complex values of $\kappa$.
This function has no zeros, and poles whenever the operator $-\partial_\tau^2 + \kappa$ has zero modes, that is, if a solution exists to the Euclidean equation of motion
\begin{equation}
 \left(-\partial_\tau^2 + \kappa \right) \, \phi(\tau) = 0 \, ,
\end{equation}
which in addition satisfies the periodicity constraint $\phi(\tau+1/T)=\phi(\tau)$. A basis of (not necessarily periodic) solutions to the equation of motion is, for any value of $\kappa$,
\begin{equation}
 \phi_\pm(\tau) = e^{\pm \sqrt{\kappa} \, \tau} .
\end{equation} 
(Wick rotated back to real time $t = -i \tau$ and for real $\kappa>0$, these become the usual oscillatory solutions to the real time equation of motion, with frequency $\sqrt{\kappa}$.)
The solutions satisfy in addition the Euclidean periodicity constraint if and only if $\pm \sqrt{\kappa} = i \w_n$, $n \in \IZ$, or equivalently if and only if $\sinh \left( \pm \frac{\sqrt{\kappa}}{2T} \right) = 0$. Thus $Z(\kappa)$ is a meromorphic function with the same zeros and poles (with the same multiplicity) as the meromorphic\footnote{Despite the square root there is no branch point at $\kappa=0$ because the function is invariant under $\sqrt{\kappa} \to -\sqrt{\kappa}$.} function
\begin{equation}
  f(\kappa) \equiv \frac{1}{\sinh^2 \bigl(\frac{\sqrt{\kappa}}{2T}\bigr)} \,.
\end{equation}
Furthermore we know\footnote{We assume this to be known in the present derivation because the analogous piece of information will be known in the black hole case: the large mass expansion of the one loop free energy is determined completely by known local expressions. In the present context it can be derived directly from the infinite product representation of the determinant by noticing that in the large $\kappa$ limit the logarithm of the product is well approximated by an integral over $n$.} that when $\kappa \to \infty$, $Z(\kappa) \to e^{-E_0/T} = e^{-\sqrt{\kappa}/T}$ while $f(\kappa) \to 4 \, e^{-\sqrt{\kappa}/T}$, where the branch of $\sqrt{\kappa}$ is taken with positive real part. Thus $\frac{4Z(\kappa)}{f(\kappa)}$ is a holomorphic function without zeros or poles which converges to 1 when $\kappa \to \infty$, that is to say, it must be equal to 1 everywhere. This reproduces (\ref{harmoscZfin}).



\subsection{Damped harmonic oscillator}

We now add dissipation to the previous example. This generalisation includes effects that are crucial in the black hole case. The partition function for the damped complex harmonic oscillator coupled to a `chemical potential' $\mu$ is up to normalization given by \cite{Weiss}
\begin{equation} \label{dampedZ}
 Z = N \, \prod_n \left((\omega_n +i \mu)^2 + 2\gamma |\omega_n| + \kappa \right)^{-1} \, , 
\end{equation}
with $\gamma>0$ and $|\mu| < m$ for stability. The normalization factor $N$ is taken to be independent of the parameters $(\kappa,\gamma,\mu)$ and fixed by requiring the correct, known, undamped $\gamma=0$ limit. The thermal frequencies $\omega_n$ are still given by (\ref{eq:bosonic}). Notice the absolute value signs around the thermal frequencies $\omega_n$ in (\ref{dampedZ}). 

The argument we used for the undamped oscillator was based on complex analyticity, and to construct a similar reasoning here, we will need to get rid of the explicit absolute value signs. We do this by splitting the product into separate parts $Z_+$, $Z_0$ and $Z_-$ containing the terms with positive, zero and negative $\w_n$ respectively, and treating these factors separately. Let us first consider the $n>0$ part, analytically continued to complex $\kappa$:  
\begin{equation}
 Z_+(\kappa) = N_+ \, \prod_{n>0} \left((\omega_n +i \mu)^2 + 2\gamma \, \omega_n + \kappa \right)^{-1} \,.
\end{equation}
This function has poles whenever a solution to the Euclidean equation of motion
\begin{equation} \label{eomplus}
 \left( -(\partial_\tau-\mu)^2 + 2 i \gamma \, \partial_\tau + \kappa \right) \, \phi(\tau) = 0 \,,
\end{equation}
satisfies the Euclidean periodicity and positive frequency constraints. A basis of solutions to the equation of motion is given by the two functions
\begin{equation}
 \phi(\tau) = e^{-z_\star \tau} \, , \qquad z_\star = \mu -i \gamma \pm \sqrt{\kappa-\gamma^2-2 i \gamma \mu} \, .
\end{equation}
(Wick rotated back to real time $t=-i\tau$, for real values of the parameters and for $\gamma > 0$, these give damped oscillatory solutions to the original real time equations of motion, with frequency ${\rm Re} \, z_\star$ and damping factor ${\rm Im} \, z_\star$.)
The above solution is periodic in Euclidean time and of positive frequency if and only if $z_\star = i \w_n$, $n=1,2,3,\ldots$, or equivalently if and only if $\Gamma\left(1+\frac{i z_\star}{2 \pi T} \right)=\infty$. Thus $Z_+(\kappa)$ is a meromorphic function of $\kappa$ with the same zeros and poles as $\prod_{z_\star(\kappa)} \Gamma\left(1+ \frac{i z_\star(\kappa)}{2 \pi T} \right)$.

The $n<0$ part of $Z$ is
\begin{equation}
 Z_-(\kappa) = N_- \, \prod_{n<0} \left((\omega_n +i \mu)^2 - 2\gamma \, \omega_n + \kappa \right)^{-1} \,.
\end{equation}
This function has poles whenever a solution to the Euclidean e.o.m.\ with $\gamma$ replaced by $-\gamma$,
\begin{equation} \label{eommin}
 \left( -(\partial_\tau-\mu)^2 - 2 i \gamma \, \partial_\tau + \kappa \right) \, \phi(\tau) = 0 \,,
\end{equation}
satisfies the Euclidean periodicity and positive frequency constraints. A basis of solutions to the equation of motion is given by the two functions\footnote{The bar notation does not mean the $\bar{z}_\star$ are obtained from the $z_\star$ by complex conjugation, although it does amount to that for real values of the parameters $\kappa,\gamma,\mu$.}
\begin{equation}
 \bar{\phi}(\tau) = e^{-\bar{z}_\star \tau} \, , \qquad \bar{z}_\star = \mu +i \gamma \pm \sqrt{\kappa-\gamma^2+2 i \gamma \mu} \, .
\end{equation}
(Wick rotated back to real time $t=-i\tau$, these give \emph{anti}-damped oscillatory solutions to the original real time equations of motion, with frequency ${\rm Re} \, \bar{z}_\star$ and damping factor ${\rm Im} \, \bar{z}_\star$.)
The above solution is periodic in Euclidean time and of positive frequency if and only if $z_\star = i \w_n$, $n=-1,-2,-3,\ldots$, or equivalently if and only if $\Gamma\left(1-\frac{i \bar{z}_\star}{2 \pi T} \right)=\infty$. Thus $Z_-(\kappa)$ is a meromorphic function of $\kappa$ with the same zeros and poles as $\prod_{\bar{z}_\star(\kappa)} \Gamma\left(1 - \frac{i \bar{z}_\star(\kappa)}{2 \pi T} \right)$.

Finally the $n=0$ part of $Z$ is 
\begin{equation}
 Z_0(\kappa) = N_0 \left(-\mu^2 + \kappa \right)^{-1} \,.
\end{equation}
This has poles whenever there is a constant solution to (\ref{eomplus}) or equivalently to (\ref{eommin}), that is whenever $\prod_{z_\star} z_\star = 0$ or equivalently $\prod_{\bar{z}_\star} \bar{z}_\star = 0$.

Putting everything together, we see that $Z(\kappa)$ is a meromorphic function with the same poles and zeros as 
\begin{equation}
 f(\kappa) \equiv \prod_{z_\star(\kappa)} \Gamma\left(1+ \frac{i z_\star(\kappa)}{2 \pi T} \right) \, \prod_{\bar{z}_\star(\kappa)} \Gamma\left(1 - \frac{i \bar{z}_\star(\kappa)}{2 \pi T} \right) \, \prod_{z_\star(\kappa)} \frac{2 \pi T}{i z_\star(\kappa)} \, .
\end{equation}
From this we can conclude that
\begin{equation} \label{ZPf}
 Z(\kappa) = e^{P(\kappa)} f(\kappa)
\end{equation}
where $P(\kappa)$ is a entire holomorphic function of $\kappa$. To determine $P(\kappa)$, we consider the $\kappa \to \infty$ limit. A direct computation gives
\begin{equation}
 \log f(\kappa) = -\frac{\sqrt{\kappa}}{T}- \frac{\gamma \log\left( \frac{(2 \pi T)^2}{\kappa} \right)}{\pi T} + \log(4 \pi^2) \, + \, {\cal O}(\kappa^{-1/2}) 
\end{equation}
On the other hand, in the $\kappa \to \infty$ limit the logarithm of the product (\ref{dampedZ}) is well approximated by replacing the sum over $n$ by an integral over $n$. Introducing a UV cutoff\footnote{The residual dependence on the UV cutoff when $\gamma \neq 0$ is related to the fact that while $\partial_\kappa \log Z$ is convergent, the same is not true for $\partial_\gamma \log Z$.} $|\omega_n|<\Lambda$ and fixing a $(\kappa,\gamma,\mu)$-independent normalization by requiring the correct $\gamma = 0$ limit, this gives
\begin{equation}
 \log Z(\kappa) = - \frac{\sqrt{\kappa}}{T}- \frac{\gamma \log\left( \frac{\Lambda^2}{\kappa} \right)}{\pi T} \, + \, {\cal O}(\kappa^{-1/2}) \, + \, {\cal O}(\Lambda^{-1}) \, .
\end{equation}
Comparing these results yields
\begin{equation}
 P(\kappa) = -\log(4 \pi^2) - \frac{2\gamma \log\left( \frac{\Lambda}{2 \pi T} \right)}{\pi T} \, .
\end{equation}
There can be no corrections in inverse powers of $\kappa$ to this expression since $P(\kappa)$ must be an entire holomorphic function, i.e.\ it has a Taylor series expansion in $\kappa$ with infinite radius of convergence.

Putting everything together, using $\Gamma(x+1)=x \Gamma(x)$ and restricting again to real values of the parameters (so the $\bar{z}_\star$ are complex conjugates of the $z_\star$) gives the final result
\begin{equation} \label{dampedharmfinal}
 Z =  e^{- \frac{2\gamma \log\left( \frac{\Lambda}{2 \pi T} \right)}{\pi T}} \, \prod_{z_\star} \frac{|z_\star|}{4\pi^2 T} \left|\Gamma\left(\frac{i z_\star}{2 \pi T} \right)\right|^2 \, .
\end{equation}
The undamped oscillator result is reproduced from this by putting $\gamma=0$ and using the identity $x \Gamma(i x) \Gamma(-i x) = \pi/\sinh(\pi x)$. 


All of these formulae generalize straightforwardly to systems with multiple quasinormal frequencies $z_\star$, by letting the products range over all quasinormal frequencies.

\subsection{Fermionic oscillator}

The partition function of the undamped fermionic oscillator is
\begin{equation}
 Z_F = \int {\mathcal{D}}\psi \, {\mathcal{D}}\bar{\psi} \, e^{-\int_0^{1/T} d\tau \, \bar{\psi}(\partial_\tau + m)\psi } \, ,
\end{equation}
and hence
\begin{equation}
 Z_F(m) = N {\det}_\text{AP}(\partial_\tau+m)
 =  N \left({\det}_\text{AP}(-\partial_\tau^2+m^2) \right)^{1/2} \, .
\end{equation}
Here the subscript `AP' indicates antiperiodic boundary conditions: $\psi(\tau+1/T)=-\psi(\tau)$. Thus, the thermal frequencies are now half integral multiples of $2 \pi T$:
\be
\omega_n = 2 \pi T (n+\half) \,, \qquad \text{(fermionic)} \,.
\ee
Apart from this fact the analysis proceeds exactly as in the bosonic case, giving instead of (\ref{harmoscZfin})
\begin{equation}
 Z_F(m) = 2 \cosh \left( \frac{m}{2T} \right) = e^{-\frac{m}{2T}} + e^{\frac{m}{2T}} \, .
\end{equation}
For multiple oscillators with mass matrix $M$ we get 
\begin{equation}
 Z_F = \prod_{z_\star} 2 \cosh \left( \frac{z_\star}{2T} \right) \,,
\end{equation}
where $z_\star$ are the fermionic normal frequencies, obtained by solving $(\partial_\tau + M)\psi = 0$ for $\psi=e^{-z_\star \tau}$.

Similarly, if we define the damped fermionic oscillator determinant as the square root of the corresponding bosonic one but with antiperiodic boundary conditions, we get instead of (\ref{dampedharmfinal})
\begin{equation} \label{dampedharmfinal2}
 Z_F = e^{ \frac{2\gamma \log\left( \frac{\Lambda}{2 \pi T} \right)}{\pi T}} \, \prod_{z_\star} \frac{2\pi}{\left|\Gamma\left(\frac{i z_\star}{2 \pi T} + \frac{1}{2} \right)\right|^2} \, .
\end{equation}

\section{Euclidean thermal geometries} \label{sec:euclthermal}

We now turn to the problem of actual interest to us, computing determinants of Laplace type operators on thermal Euclidean spaces obtained by Wick rotating static spacetimes. 

\subsection{Asymptotically AdS black holes}

Consider a static asymptotically AdS spacetime with a horizon. A specific example is Schwarzschild-AdS$_{d+1}$:
\begin{equation} \label{generalAdSmetric}
 ds^2 = - V(r) \, d t^2 + \frac{1}{V(r)} \, dr^2 + r^2 \, d \Omega_{d-1}^2 \, , 
\end{equation}
where $V(r)=1-\frac{m}{r^{d-2}}+\frac{r^2}{L^2}$. However the arguments which we will give in the following are not restricted to this example, and apply to arbitrary static backgrounds. Geometries with horizons are the most interesting as computing determinants in these cases will require treating the positive and negative frequency modes separately, similarly to the damped oscillator.

To compute the one loop partition function at temperature $T$, we Wick rotate and identify:
\begin{equation}\label{period}
 t = - i\tau \, , \qquad \tau \simeq \tau + 1/T \, ,
\end{equation}
so the metric becomes Euclidean and periodic in the Euclidean time $\tau$. The fact that the spacetime is static means that the metric components are independent of $\tau$ and that there are no space-time mixing terms. The horizon of the original black hole becomes the origin of the Wick rotated Euclidean space \cite{Gibbons:1976ue}. The time circle contracts to a point there. For a suitable radial coordinate $\rho$, the metric near the origin is of the standard flat form 
\begin{equation} \label{rhotheta}
 ds^2 \approx d\rho^2 + \rho^2 d\theta^2 + ds_\perp^2 \, , \qquad 
 \theta \equiv 2 \pi T \tau \, ,
\end{equation} 
where $ds_\perp^2$ is the transversal metric at the horizon. Regularity of this metric at $\rho=0$ fixes the periodicity of the angular coordinate $\tau$ to be (\ref{period}), and hence the black hole temperature to be $T$. 

For another suitable radial coordinate $z$, the metric near the AdS boundary $z=0$ is of the form
\begin{equation}
 ds^2 \approx \frac{L^2}{z^2} \left( dz^2 + d\tau^2 \right) + ds_\perp^2 \, ,
\end{equation}
where $L$ is the AdS radius.

For concreteness, say that we want to compute the contribution to the one loop partition function in such a background from a massive complex scalar $\phi$. This is given by 
\begin{equation} \label{Zdefscalar}
 Z(\Delta) = \int {{\cal D} \phi} \, e^{-\int \phi^* \left( -\nabla^2 + m^2 \right) \phi} \, \propto \, \frac{1}{\det\left( -\nabla^2+m^2 \right)} \, ,
\end{equation}
defined with the $z \to 0$ boundary condition 
\begin{equation} \label{boundarycond}
 \phi \sim z^\Delta \left(1+{\cal O}(z^2) \right) \, ,
\end{equation}
where $\Delta$ is a root of $\Delta(\Delta-d)=(mL)^2$. With this asymptotic behavior, the action is positive definite for real $\Delta$ and finite for $\Delta > \frac{d-2}{2}$, which is the unitarity bound of the dual CFT \cite{Klebanov:1999tb}. The fact that the boundary condition (\ref{boundarycond}) is specified in terms of $\Delta$ rather than the mass $m^2$ will shortly motivative us to use analyticity arguments in $\Delta$ rather than $m^2$.

We also require regularity at the origin $\rho = 0$. In terms of the complex coordinate 
$u \equiv \rho \, e^{-i \theta}$ with $\rho$ and $\theta$ as in (\ref{rhotheta}), regularity means that $\phi$ should have a regular Taylor series expansion in $u$ and $\bar{u}$ about the origin. For a mode with thermal frequency quantum number $n$, we therefore have the $\rho \to 0$ behavior
\begin{equation} \label{ndependence}
 \phi \sim
\left\{  \begin{array}{lr}
 u^n = \rho^n e^{- i n \theta} = \rho^{\omega_n/2 \pi T} e^{-i \omega_n \tau} & \qquad \mbox{if} \quad n \geq 0 \, , \\
 \bar{u}^{-n} = \rho^{-n} e^{- i n \theta} = \rho^{-\omega_n/2 \pi T} e^{-i \omega_n \tau} & \qquad \mbox{if} \quad n \leq 0 \, ,
\end{array} \right.
\end{equation}
where the $\omega_n$ are the bosonic thermal frequencies as in (\ref{eq:bosonic}).


Paralleling our warmup examples, we will determine $Z(\Delta)$ by analytically continuing to complex $\Delta$ and matching poles and zeros. Because of the nonanalytic dependence on $n$ exhibited in (\ref{ndependence}), we will be forced to split the partition function into positive and negative frequency parts, as we had to do for the damped harmonic oscillator. In fact much of the following analysis will mimic the analysis done for the damped harmonic oscillator. Let us start with the positive frequency part $Z_+(\Delta)$, defined 
by (\ref{Zdefscalar}) restricted to functions with only positive frequency Fourier coefficients. This has no zeros, and poles on complex values of $\Delta$ for which a zeromode exists, that is to say, a solution to the equation of motion\footnote{The equation of motion near $\rho=0$ implies $\phi_\star \sim \rho^{\mp i z_\star/2 \pi T} \, e^{-z_\star \tau}$. From (\ref{ndependence}) it follows that positive frequency solutions are on the $\rho^{- i z_\star/2 \pi T}$ branch, while negative frequency solutions are on the $\rho^{+ i z_\star/2 \pi T}$ branch.} 
\begin{equation}  \label{QNMEOM}
 \left( -\nabla^2 + \mbox{$\frac{\Delta(\Delta-d)}{L^2}$} \right) \, \phi_\star = 0 \, ,
 \qquad \phi_\star \sim \rho^{-i z_\star/2 \pi T} \, e^{-z_\star \tau} \quad (\rho \to 0) \, ,
\end{equation}
satisfying the boundary condition (\ref{boundarycond}) at infinity, which in addition is periodic in $\tau$. The latter is equivalent to requiring\footnote{It is important to keep in mind that we are considering complex $\Delta$ here. For real $\Delta$, there will generically be no zeromodes; the poles of $Z(\Delta)$ are off the real axis.}
$z_\star(\Delta) = i \omega_n$, $n=1,2,3,\ldots$, or equivalently $\Gamma\left(1+\frac{i z_\star}{2 \pi T} \right)= \infty$. 

When we Wick rotate the solutions to real time $t=-i\tau$, the $z_\star$ appear as complex frequencies of modes with near-horizon behavior 
\begin{equation}
 \phi_\star \propto \exp \left( - i z_\star (x+t) \right) \, , \qquad x \equiv \frac{\ln \rho}{2 \pi T} \, .
\end{equation} 
The horizon is at $x = -\infty$, hence these modes satisfy so-called \emph{ingoing} boundary conditions at the horizon as well as $\phi \sim z^\Delta$ boundary conditions at infinity. For physical, real values of $\Delta$, this precisely defines the standard quasinormal modes of the AdS black hole \cite{Berti:2009kk}. Thus we arrive at the important conclusion that for the physical, real values of $\Delta$, \emph{the $z_\star$ are the black hole quasinormal frequencies}.

Stability of the background requires that all quasinormal frequencies have negative imaginary part; if not, some modes will be exponentially growing with time. 

For the negative frequency part of the determinant we follow a similar reasoning, except that now we need to select the other branch of solutions to the equations of motion. The poles in $Z_-(\Delta)$ are associated to solutions to the equation of motion $\bar{\phi}_\star$ with $\rho \to 0$ behavior $\sim \rho^{+i \bar{z}_\star/2 \pi T} \, e^{-\bar{z}_\star \tau}$, for which $\bar{z}_\star = i \omega_n$, $n=-1,-2,-3,\ldots$, or equivalently $\Gamma\left(1- \frac{i \bar z_\star}{2 \pi T} \right)= \infty$. 

After Wick rotating to real time, these solutions satisfy \emph{outgoing} boundary conditions at the black hole horizon, and the $\bar{z}_\star$ are the `anti'-quasinormal frequencies. Despite what the notation might suggest, the sets of these anti-quasinormal modes and -frequencies are not necessarily complex conjugates of the standard ingoing quasinormal modes. However for real values of the parameters, if the operator $\nabla^2$ is PT invariant (as is often the case; for example it is true for all examples we will consider in this paper, as well as for the planar dyonic black hole in four dimensions we considered in \cite{us}), these outgoing modes can be obtained from the ingoing modes by acting with the PT symmetry, and then in particular the set of ingoing and outgoing quasinormal frequencies are each others complex conjugate. 

The $n=0$ contribution can be treated similarly.

Putting everything together,\footnote{As with the damped harmonic oscillator for the $n=0$ part we can work either with the $z_\star$ or with the $\bar{z}_\star$, and we have $\prod_{z_\star} z_\star = \prod_{\bar{z}_\star} \bar{z}_\star  = \prod_{z_\star,\bar{z}_\star} \sqrt{z_\star \bar{z}_\star} $.}, we see that $Z(\Delta)$ is a meromorphic function of $\Delta$ with the same poles and zeros as 
the function $\prod_{z_\star,\bar{z}_\star} \sqrt{z_\star \bar{z}_\star} \, \Gamma\left(\frac{i z_\star}{2 \pi T} \right) \Gamma\left(\frac{-i \bar{z}_\star}{2 \pi T} \right)$. Therefore 
\begin{equation} \label{QNMfinal}
 Z_B = e^{\text{P} (\Delta)} \, \prod_{z_\star,\bar{z}_\star} \frac{\sqrt{z_\star \bar{z}_\star}}{4 \pi^2 T} \, \Gamma\left(\frac{i z_\star}{2 \pi T} \right) \Gamma\left(\frac{-i \bar{z}_\star}{2 \pi T} \right) \, ,
\end{equation}
$P(\Delta)$ is a nonsingular holomorphic function of $\Delta$, which we will later determine by matching the $\Delta \to \infty$ behavior. The function $P(\Delta)$ is analogous to $P(\kappa)$ in (\ref{ZPf}).
To make this more precise, we have to address regularization and renormalization, as both the determinant (\ref{Zdefscalar}) and the product (\ref{QNMfinal}) are in fact UV divergent. We will postpone this to the next section, where we will consider concrete examples of how this works in practice. We will see that $P(\Delta)$ is in fact a \emph{polynomial} in $\Delta$, determined by computing a finite number of integrals over spacetime of local curvature invariants. To emphasize the polynomial dependence we will write $P(\Delta) =  {\rm Pol}(\Delta)$. 

As we mentioned, in many cases $z_\star$ and $\bar z_\star$ are complex conjugates. In this case, the above expression reduces to (\ref{eq:statement}).

An alternative and sometimes more useful way of writing this formula is
\begin{equation} \label{QNMfinalprodform}
 Z_B = e^{\text{Pol}(\Delta)} \, \prod_{z_\star} \frac{\sqrt{z_\star \bar{z}_\star}}{2\pi T} \, \prod_{n \geq 0} \left(n+\frac{i z_\star}{2 \pi T} \right)^{-1} \left(n - \frac{i \bar{z}_\star}{2 \pi T} \right)^{-1} \, ,
\end{equation}
where the divergent products over $n$ are understood to be regularized, as is the product over the $z_\star$ (more on this in the next section).

Although we assumed a single complex scalar in a black hole background, exactly the same formula holds for different numbers of bosonic fields. For fermions we get, similar to (\ref{dampedharmfinal2}):
\begin{equation} \label{fermifinal}
 Z_F = e^{\text{Pol}(\Delta)} \prod_{z_\star} \frac{2\pi}{\Gamma\left(\frac{1}{2} + \frac{i z_\star}{2 \pi T} \right) \Gamma\left(\frac{1}{2} - \frac{i \bar{z}_\star}{2 \pi T} \right) } \, .
\end{equation}
or, in product form
\begin{equation}
 Z_F = e^{\text{Pol}(\Delta)} \prod_{z_\star} \prod_{n\geq 0} \left(n+\frac{1}{2} + \frac{i z_\star}{2 \pi T} \right) \left(n+ \frac{1}{2} - \frac{i \bar{z}_\star}{2 \pi T} \right) \, .
\end{equation}
Here the $z_\star$ are again the quasinormal frequencies, obtained by solving the appropriate Dirac equation of motion.

\subsection{Thermal AdS}

Geometries without a horizon have less structure; the quasinormal modes become normal modes, and our formalism is correspondingly less useful. However, we begin with such simpler examples as they will allow a transparent introduction to the computation of the $e^{\text{Pol}(\Delta)}$ terms later. Furthermore, even in these cases our approach appears to have some technical advantages compared to previous computations.

The metric of $AdS_{d+1}$ in global coordinates is of the form (\ref{generalAdSmetric}) 
with $V(r)=1 + \frac{r^2}{L^2}$. The same arguments as for the black hole may be applied to this case. The difference is that the thermal circle now nowhere collapse to a point, so we are free to (but of course don't have to) treat positive and negative frequencies in one go, like we could for the harmonic oscillator in the absence of damping. Exactly the same formulae as for the black hole spacetimes continue to hold, although since there is no horizon in the Lorentzian geometry, the Lorentzian interpretation as ingoing and and outgoing modes is lost.  Because the frequencies $z_\star$ are real now (as there is no dissipation), we can bring the formulae to a form similar to those of the undamped harmonic oscillator using $x \Gamma(i x) \Gamma(-i x) = \pi/\sinh(\pi x)$. In addition, the frequencies have a direct interpretation as one particle energies (with respect to global AdS time): $|z_\star|=E_\star$. Thus, for a complex boson:
\begin{equation}
 Z_B = e^{{\rm Pol}(\Delta)} \, \prod_{z_\star} \frac{1}{2 \sinh \frac{|z_\star|}{2 T}} = e^{{\rm Pol}(\Delta)}
 \prod_{z_\star} \frac{e^{-\frac{|z_\star|}{2T}}}{1-e^{-\frac{|z_\star|}{T}}} \, ,
\end{equation}
which we recognize as a free multi-particle partition function $Z=\Tr \, e^{-H/T}$. For fermions:
\begin{equation}
 Z_F = e^{{\rm Pol}(\Delta)} \, \prod_{z_\star} 2 \cosh |z_\star| = e^{{\rm Pol}(\Delta)} \, \prod_{z_\star} e^{\frac{|z_\star|}{2 T}} \left( 1 + e^{\frac{-|z_\star|}{T}} \right) \,,
\end{equation}
which again we recognize as $\Tr \, e^{-H/T}$.

The exact normal frequencies of global AdS$_{d+1}$ are well known. For scalars \cite{Aharony:1999ti}:
\begin{equation} \label{AdSnormal}
 z_{n,\ell,\pm} = \pm \frac{2n + \ell + \Delta}{L}  \, , \qquad n,\ell = 0,1,2,\ldots \,,
\end{equation}
where $L$ is the AdS radius, $\ell$ is the angular momentum quantum number, and $\Delta$ the conformal dimension, related to the mass by $(m L)^2 = \Delta(\Delta-d)$. The degeneracy $D_\ell$ of each frequency equals the degeneracy of the $\ell$th angular momentum eigenvalue on $S^{d-1}$, i.e.\
\begin{equation} \label{Dell}
 D^{(d-1)}_{\ell}={\ell+d-1 \choose d-1} - {\ell+d-3 \choose d-1} = \frac{2 \ell + d -2}{d-2} {\ell+d-3 \choose d-3} \, .
\end{equation} 
For example when $d=3$, this gives $D_\ell=2\ell+1$. Somewhat degenerate cases are $d=2$, for which one should take $D_0 = 1$ and $D_\ell = 2$ for $\ell > 0$, and $d=1$, for which $D_0=D_1=1$ and $D_\ell=0$ for $\ell>1$. The complex boson one loop partition function is then
\begin{equation} \label{thermalAdSboson}
 Z_B = e^{{\rm Pol}(\Delta) - \sum_{\ell,n} 2D_\ell \frac{2n + \ell + \Delta}{2LT}}  \, \prod_{n,\ell} {\left(1-e^{-\frac{2n + \ell + \Delta}{LT}}\right)^{-2D_\ell}} \, .
\end{equation}
The sum in the first exponential is clearly UV divergent. This is the usual boson zero point energy divergence and can be absorbed in the UV counterterms in ${\rm Pol}(\Delta)$. We will return to this in the next section, where we will explicitly compute the full AdS$_3$ partition function.

\subsection{de Sitter space} \label{sec:dS1}

The above arguments are not restricted to asymptotically AdS spaces. They apply equally well to for example asymptotically de Sitter spaces.\footnote{We thank Dionysis Anninos for his encouragement.} Let us consider pure $dS_{d+1}$ to illustrate this. Static patches of this spacetime have horizons that are formally similar to black hole horizons \cite{Gibbons:1977mu}.

The de Sitter metric in static coordinates is of the form (\ref{generalAdSmetric}) with 
$V(r) = 1 - \frac{r^2}{L^2}$. It is formally obtained from the AdS case by substituting $L \to iL$. A causal diamond is given by the coordinate patch $0 \leq r \leq L$. The (cosmological) horizon is at $r=L$. After Wick rotation the causal diamond becomes a sphere. The dS temperature is easily read off as $T=\frac{1}{2 \pi L}$, by computing the radius of the $\tau$ circle required for regularity.

The Lorentzian quasinormal modes $\phi_\star$ have ingoing boundary conditions at the horizon. As for black hole horizons, the Wick rotation of this statement is simply regularity for positive frequencies. Therefore this is again the relevant boundary condition for the purpose of computing determinants.

The de Sitter quasinormal frequencies for a scalar are \cite{LopezOrtega:2006my}
\begin{equation} \label{dSQNM}
 z_{n,\ell,\pm} = -i \frac{2n + \ell + \Delta_\pm}{L}  \, , \qquad n,\ell = 0,1,2,\ldots \,,
\end{equation}
and $\bar{z}_{n,\ell,\pm} = (z_{n,\ell,\pm})^*$, where the notation parallels that in (\ref{AdSnormal}), except that $\Delta_\pm$ are now roots of $\Delta(\Delta-d)=-(m L)^2$, so in particular for $m>d/2$, the $\Delta_\pm$ become complex. In AdS, the boundary conditions fixed the root $\Delta$, but because now there is no boundary, both $\Delta_+$ and $\Delta_-$ appear. Inserting these frequencies in (\ref{QNMfinal}) or (\ref{QNMfinalprodform}) and using the degeneracies $D_\ell$ from (\ref{Dell}) then produces the one loop thermal partition function for a complex scalar in dS$_{d+1}$, or equivalently the inverse determinant of the massive Laplacian on the sphere $S^{d+1}$. We will do this explicitly in a later section.

\section{Local contributions and examples} \label{sec:locex}

So far we have ignored UV divergences and renormalization, and have hidden them in the as yet unspecified term ${\rm Pol}(\Delta)$, which we claimed is a polynomial in $\Delta$ and obtained from the computation of a finite number of local terms. We now turn to these issues. To do this, we will study a number of concrete examples, which at the same time will serve to test the above expressions, by reproducing some previously derived results.

\subsection{Thermal AdS$_3$}

We start with the thermal AdS partition function (\ref{thermalAdSboson}). For simplicity we will focus on AdS$_3$, which has a metric of the form (\ref{generalAdSmetric}) with $V(r) = 1 + \frac{r^2}{L^2}$, $d\Omega_1 = d\phi$,
with $\phi \simeq \phi + 2 \pi$. We will consider a real boson instead of a complex one, which means we should take the square root of (\ref{thermalAdSboson}). We may combine the products over $\ell,n$ with multiplicity $D_\ell$ into a single product over $\kappa=2n+\ell$ with multiplicity $\kappa+1$:
\begin{equation} 
 \log Z = {\rm Pol}(\Delta) - \sum_{\kappa=0}^\infty (\kappa+1) \frac{\kappa + \Delta}{2LT}  \, - \,\sum_{\kappa=0}^\infty (\kappa+1) \log \left(1-e^{-\frac{\kappa + \Delta}{LT}}\right)\, . 
\end{equation} 
The last sum converges, but the first one is UV divergent --- this is the usual zero point energy divergence of a free boson. We can regulate this sum for example by a simple energy cutoff. Recall that ${\rm Pol}(\Delta)$ so far has only been specified to be a polynomial in $\Delta$ which contains UV counterterms as well as terms necessary to give the correct large $\Delta$ behavior. We may therefore absorb any UV divergence which depends \emph{polynomially} on $\Delta$ into ${\rm Pol}(\Delta)$, and write:
\begin{equation} \label{thermalAdSlogZ}
 \log Z = {\rm Pol}(\Delta) \, - \,\sum_{\kappa} (\kappa+1) \log \left(1-e^{-\frac{\kappa + \Delta}{LT}}\right)\, . 
\end{equation} 
The second sum agrees precisely with the finite, nonlocal part of the scalar one loop partition function obtained for thermal AdS in \cite{Giombi:2008vd} 
by explicitly solving the heat equation and applying the method of images.

The polynomial ${\rm Pol}(\Delta)$ is fixed by requiring the correct large $\Delta$ behavior. More precisely, since the second term in (\ref{thermalAdSlogZ}) vanishes when $\Delta \to \infty$, we need
\begin{equation} \label{ZPrel}
 \lim_{\Delta \to \infty} \left( \log Z(\Delta) - {\rm Pol}(\Delta) \right) = 0 \, . 
\end{equation}
To find ${\rm Pol}(\Delta)$ we therefore need an independent way of determining the large $\Delta$ behavior of $\log Z(\Delta)$. Because at large $\Delta$ the scalar becomes very massive, this will be determined by purely local expressions. These can be found straightforwardly from the heat kernel coefficients of the Laplacian. In the next part we briefly recall this technology (for a review, see e.g. \cite{Vassilevich:2003xt}).

\subsection{Local terms from heat kernel coefficients}

We may write  
\begin{equation}
 \log \left( \lambda/\lambda_0 \right) = - \int_0^\infty \frac{dt}{t} \left( e^{-\lambda t} - e^{-\lambda_0 t} \right) \, ,
\end{equation}
for some fixed reference $\lambda_0$, as can be seen by differentiating both sides with respect to $\lambda$. From this observation, we get the following formal heat kernel representation of the determinant of an operator $D+m^2$:
\begin{equation}
 \log \det (D+m^2) = \Tr \log (D+m^2) = {\rm const.} - \int_0^\infty \frac{dt}{t} \, \Tr \, e^{-t (D+m^2)} \,  .
\end{equation}
Here and below ``const.'' should be thought of as some fixed reference determinant.

For any operator $D$ in $n=d+1$ dimensions of the form $D=-(\nabla^2+E)$, with arbitrary gauge and metric covariant derivative $\nabla$ and background field $E$, one has the following $t \to 0$ asymptotic expansion of the trace of the heat kernel:
\begin{equation}
 {\rm Tr} \, e^{-t D} \simeq (4 \pi t)^{-\frac{n}{2}} \sum_{k=0}^\infty a_k(D) \, t^{k/2} \, .
\end{equation}
The heat kernel coefficients $a_k$ have \emph{universal} expressions in terms of local curvature invariants. If we ignore boundary terms, then $a_k = 0$ for odd $k$ and for even $k$ we have $a_k(D) = \int d^{n} x \, \sqrt{g} \, \tr \, a_k(D,x)$, with \cite{Vassilevich:2003xt}:
\begin{eqnarray}
 a_0(D,x) &=& 1 \,,  \\
 a_2(D,x) &=& \frac{1}{6} \left( 6 \, E + R \right) \,, \\
 a_4(D,x) &=& \frac{1}{360} \left( 
 60 \, \nabla^2 E + 60 \, R E + 180 \, E^2 
 + 12 \, \nabla^2 R + 5 \, R^2 \right. \\
 && \left. - 2 \, R_{\mu\nu} R^{\mu\nu} 
 +2 \, R_{\mu\nu\rho\sigma} R^{\mu\nu\rho\sigma} + 30 \, F_{\mu\nu} F^{\mu\nu} \, \right) . 
\end{eqnarray}
General expressions are known up to $a_{10}$. 

From this asymptotic expansion we can extract the $m \to \infty$ behavior of the determinant, which only depends on the heat kernel coefficients $a_0$ to $a_n$:
\begin{equation} \label{largemassexp}
 \log \det(D+m^2) \, = \, {\rm const.} - (4 \pi)^{-\frac{n}{2}} \sum_{k=0}^n a_k(D) \, \int_0^\infty \frac{dt}{t} \,  t^{\frac{k-n}{2}} \, e^{-t m^2} \, + \, {\cal O}(m^{-1}) \, .
\end{equation}
Indeed, the higher order contributions give finite integrals which manifestly vanish when $m \to \infty$. Furthermore, all UV divergent terms are contained in this truncated series.

The determinant can be regularized as
\begin{equation}
 \log {\det}_\Lambda (D+m^2) = {\rm const.}_{\Lambda} -\int_0^\infty \frac{dt}{t} \, f(\Lambda^2 t) \, {\rm Tr} \, e^{-t (D+m^2)} \,,
\end{equation}
where $f(x)$ is some function which goes to zero sufficiently fast when $x \to 0$ and approaches 1 when $x \to \infty$. Therefore $f(\Lambda^2 t)=1$ except infinitesimally close to $t=0$ when $\Lambda \to \infty$. Popular choices are $f(x) = \Theta(x-1)$ with $\Theta$ the Heaviside step function, or $f(x) = \left(1-e^{-x} \right)^k$ with $k>\frac{n}{2}$.   
Using the step function, for example in dimension $n=3$, this then gives for the regularized version of (\ref{largemassexp})
\begin{eqnarray} \label{largemass3d}
 \log {\det}_\Lambda(D+m^2) &=& {\rm const.}_{\Lambda} \, + \, \frac{1}{(4 \pi)^{3/2}} \left( - \frac{2 \Lambda^3}{3} + 2 m^2 \Lambda - \frac{4 \sqrt{\pi}}{3} m^3 \right) \cdot a_0(D) \nonumber \\ 
 && +
 \frac{1}{(4 \pi)^{3/2}} \left( -2 \Lambda + 2 m \sqrt{\pi} \right) \cdot a_2(D) \,\, + \,{\cal O}(m^{-1}) \, .
\end{eqnarray}
Here $a_0=\int d^3x \, \sqrt{g}$ and for a scalar $E=0$, so $a_2 = \int d^3 x \, \sqrt{g} \, \frac{1}{6} R$. These terms simply renormalize the cosmological constant and Newton's constant. In dimensions $n=4,5$ the couplings associated to terms in $a_4(D)$ will also be renormalized, and so on.

\subsection{Result for thermal AdS$_3$}

Let us now use the result of the previous subsection to determine ${\rm Pol}(\Delta)$ for the thermal AdS$_3$ example, according to (\ref{ZPrel}). In AdS$_3$ we have $R=-6/L^2$. Using $\log Z={\rm const.} -\log \det(D+m^2)$, $m^2=\Delta(\Delta-2)/L^2$, and absorbing the ($\Delta$-independent) $\Lambda^3$ term in the constant, (\ref{largemass3d}) boils down to the simple expression:
\begin{equation}
 \log Z(\Delta) = {\rm const.} + \int d^3x \sqrt{g} \, \left( \frac{(\Delta-1)^3}{12 \pi L^3} - \frac{(\Delta-1)^2\Lambda}{8 \pi^{3/2}L^2} 
 \right) 
 \, + {\cal O}(\Delta^{-1}) \, .
\end{equation}
Equation (\ref{ZPrel}) then fixes, in this regularization scheme, ${\rm Pol}(\Delta)$ to be exactly the polynomial part of this expression. In conclusion, we get our final expression for the \emph{exact} one loop partition function for a neutral scalar in thermal AdS$_3$ in this regularization scheme:
\begin{equation} \label{finalAdSZ}
 \log Z = {\rm const.} + \int d^3x \sqrt{g} \, \left( \frac{(\Delta-1)^3}{12 \pi L^3} - \frac{(\Delta-1)^2\Lambda}{8 \pi^{3/2}L^2} 
 \right) \,  \,
+ \, \, \log \prod_\kappa \left(1-q^{\kappa + \Delta}\right)^{-(\kappa+1)}\, ,
\end{equation}
where 
\begin{equation} \label{modtaudef}
 q = e^{2 \pi i \tau} \, , \qquad \tau = \frac{i}{2 \pi L T} \, ,
\end{equation}
and ``const.'' indicates that the overall normalization of the partition function has not been fixed; in other words the above should be thought of as an expression for $\log Z/Z_0$ for some fixed reference $Z_0$.

The UV finite local term does not depend on the regularization scheme and agrees with the finite local term found in \cite{Giombi:2008vd}. Note that this is nontrivial: we determined this term by matching the large $\Delta$ asymptotics, and the analyticity arguments which we used to derive our determinant formula then fixes this to be the exact result valid for any value of $\Delta$. Note that the analyticity argument does not hold for $m$ because, as noted below (\ref{boundarycond}), the boundary conditions are not analytic in $m$, but rather in $\Delta=1 \pm \sqrt{(mL)^2+1}$. And indeed the above expression (\ref{finalAdSZ}) has an infinite series of $1/m$ corrections to the leading large $m$ result.

For scalars with $-1 \leq m^2 \leq 0$, we can impose either $\Delta=\Delta_+ = 1+\sqrt{(mL)^2+1}$ or $\Delta=\Delta_-=1-\sqrt{(mL)^2+1}$ boundary conditions \cite{Klebanov:1999tb}. In the CFT dual, going from $\Delta_-$ to $\Delta_+$ boundary conditions corresponds to turning on a double trace deformation of the UV CFT \cite{Witten:2001ua} in which the operator dual to $\phi$ has conformal dimension $\Delta_-$, causing the theory to flow to a new IR fixed point in which the operator dual to $\phi$ acquires conformal dimension $\Delta_+$. The central charges of the two theories are different. In the bulk this translates to different one loop vacuum energy densities $V$. This difference is obtained by subtracting $\log Z(\Delta_-)$ from $\log Z(\Delta_+)$. In the $T \to 0$ limit:
\begin{equation}
 V(\Delta_+) - V(\Delta_-) =  -\frac{(\Delta_+-1)^3}{12 \pi L^3} + \frac{(\Delta_--1)^3}{12 \pi L^3} = - \frac{(\Delta_+-1)^3}{6 \pi L^3} \, ,
\end{equation}
reproducing the results of \cite{Gubser:2002zh,Hartman:2006dy}. Note that this is finite and independent of the regularization. Since the both the difference and the cosmological constant are negative, the $\Delta_+$ theory has a smaller central charge than the $\Delta_-$ theory, as expected.

\subsection{The BTZ black hole}

We now turn to actual black hole spacetimes. The simplest example is the BTZ black hole \cite{Banados:1992wn}, for which an exact computation of the one loop partition function is possible using our formula, which can be compared again to \cite{Giombi:2008vd}. This provides another test of our formula. The one loop free energy of a scalar in a BTZ background was first computed in \cite{Mann:1996ze}.

The BTZ black hole metric with zero angular momentum is of the form (\ref{generalAdSmetric}) with  $V(r) = -M + \frac{r^2}{L^2}$, $M>0$, $d\Omega_1 = d\phi$, $\phi \simeq \phi + 2 \pi$. (Note that the thermal AdS$_3$ metric is obtained by putting $M \equiv -1$.) Regularity at the origin $r=r_+=L\sqrt{M}$ requires $2 \pi T = \frac{1}{L \sqrt{M}}$. 

The quasinormal frequencies of a neutral massive scalar of mass $m^2=\Delta(\Delta-1)/L^2$ in a non-rotating BTZ black hole background are \cite{Cardoso:2001hn,Birmingham:2001pj,Berti:2009kk}
\begin{eqnarray}
 z_{p,s,\pm} &=& \pm \frac{p}{L} - 2 \pi T i (\Delta + 2 s) \, , \qquad s=0,1,2,\ldots, \quad
 p=0,\pm1,\pm2,\ldots  
\end{eqnarray}
and $\bar{z}_{p,s,\pm} = ({z}_{p,s,\pm})^*$. Here $p$ is the momentum quantum number along the spatial circle and $\pm$ distinguishes left and right movers. Note that each $p$ therefore appears twice. Equation (\ref{QNMfinalprodform}) (with a factor 1/2 because the scalar is real) becomes
\begin{eqnarray}
 -\log Z &=& -{\rm Pol}(\Delta) + \sum_{s\geq 0,n>0,p} \left[ \log \left(n+2s+\Delta + i \frac{p}{2 \pi L T} \right) + \log \left(n+2s+\Delta - i \frac{p}{2 \pi L T} \right) \right] \nonumber \\
 &&+\frac{1}{2} \sum_{s\geq0,p} \left[ \log \left(2s+\Delta + i \frac{p}{2 \pi L T} \right) + \log \left(2s+\Delta - i \frac{p}{2 \pi L T} \right) \right] \\
 &=& -{\rm Pol}(\Delta) + \frac{1}{2} \sum_{\kappa \geq 0,p} (\kappa+1) \log \left((\kappa+\Delta)^2 + \left(\frac{p}{2 \pi L T}\right)^2 \right) \, ,
\end{eqnarray}
where we combined the sums over $n$ and $s$ into a sum over $\kappa=0,1,2,\ldots$ with multiplicity $\kappa+1$, and used $\log(x+i y) + \log(x-i y)=\log(x^2+y^2)$. Extracting divergent sums while absorbing as before all terms polynomial in $\Delta$ into ${\rm Pol}(\Delta)$, and using the identity $\sum_{p\geq 1} \log \left(1+ \frac{x^2}{p^2} \right) = \log \frac{\sinh \pi x}{\pi x} = \pi x - \log(\pi x) + \log \left( 1 - e^{-2 \pi x} \right)$, we get
\begin{equation} \label{logZformulaaaa}
 \log Z = {\rm Pol}(\Delta) + \log \prod_{\kappa \geq 0} \left(1 - \tilde{q}^{\kappa+ \Delta} \right)^{-(\kappa+1)} \, .
\end{equation}
where
\begin{equation}\label{eq:qtilde}
 \tilde{q} = e^{2 \pi i \tilde{\tau}} \, , \qquad \tilde{\tau} = 2 \pi i L T = -\frac{1}{\tau} \, ,
\end{equation}
where $\tau$ was defined before in (\ref{modtaudef}). Since again the second term in (\ref{logZformulaaaa}) goes to zero when $\Delta \to \infty$, and BTZ is locally identical to AdS$_3$, the integrand of the local term ${\rm Pol}(\Delta)$ will be exactly as in the thermal AdS$_3$ case (\ref{finalAdSZ}). 

The product in (\ref{logZformulaaaa}) is related by a modular transformation $\tau \to -1/\tau$ to the product in (\ref{finalAdSZ}), in agreement with \cite{Giombi:2008vd} and as expected on general grounds. This provides a nontrivial check of the quasinormal mode formula for the partition function.

Before moving on we make a brief comment about holography at a quantum level.
Notice that $L T \to 0$ in (\ref{eq:qtilde}) implies $\tilde{q} \to 1$ causing $\log Z_{\rm BTZ}$ to diverge as $1/(LT)^2$. The classical bulk action in contrast scales $\propto LT$, so at sufficiently small $T$, the one loop correction becomes larger than classical saddle point contributions and we lose control of the loop expansion. Furthermore, the divergent scaling of the correction violates extensivity of the dual boundary theory free energy.
However, well before this happens, the thermal AdS saddle point comes to dominate the path integral --- this is the Hawking-Page phase transition \cite{Hawking:1982dh}. Similar remarks hold for the large $L T$ limit of $Z_{\rm AdS_3}$. Quantum gravity comes to the rescue of holography, as one might have anticipated.

The above results can be generalized to arbitrary quotients of AdS$_{d+1}$.\footnote{Exact results for such quotients of AdS$_{d+1}$ were also obtained in \cite{diaz}, using a different (`holographic') approach of which some elements are somewhat reminiscent of elements in our discussion.}

\subsection{dS$_2$} \label{sec:dS2}

In even dimensions the heat kernel large $m$ expansion (\ref{largemassexp}) involves terms logarithmic in $m$. One might worry that this would produce $\log \Delta$ terms and that large $\Delta$ matching thus would force the inclusion of $\log \Delta$ terms in $\text{Pol}(\Delta)$, invalidating our analyticity arguments. To illustrate that this is not so, and as another check of our general reasoning, we now consider the case of a scalar in two dimensional de Sitter space (which Wick rotates to $S^2$) in some detail.  We will consider general $dS_{d+1}$ in the following section.

As reviewed in section (\ref{sec:dS1}) the dS$_2$ massive scalar quasinormal frequencies are
\begin{equation}\label{bababam}
 z_{\kappa,\pm} = -i \, \frac{\kappa + \Delta_\pm}{L} \, , \qquad
 \Delta_\pm = \frac{1}{2} \pm i \nu  \, ,
 \qquad \nu \equiv \sqrt{(mL)^2-1/4} \, ,
\end{equation}
with $\kappa=0,1,2,\ldots$ and degeneracy 1. This can be obtained from (\ref{dSQNM}) by combining $\kappa=2n+\ell$ for $\ell=0,1$. Substituting into (\ref{QNMfinalprodform}) with $T=\frac{1}{2 \pi L}$ produces:
\begin{eqnarray}
 \log Z \, - {\rm Pol}(\nu) &=& - 2 \sum_{\k,n \geq 0,\pm} \log\left( \k + \Delta_\pm \right) 
 + \sum_{\k \geq 0,\pm} \log\left( \k + n + \Delta_\pm \right) \\
 &=& - \sum_{\pm,r \geq 0} (2 r + 1) \log \left( r + \Delta_{\pm} \right) \label{sumr} \\
 &=& \sum_{\pm} 2 \, \zeta'(-1,\Delta_\pm) - (2 \Delta_\pm - 1) \, \zeta'(0,\Delta_\pm) \, .
\end{eqnarray}
Here we zeta function regularized the series; $\zeta$ is the Hurwitz zeta function defined by analytic continuation of $\zeta(s,x) = \sum_{n \geq 0} (x+n)^{-s}$, and we denote $\zeta'(s,x) \equiv \partial_s \zeta(s,x)$. The `Pol' term should be understood as a polynomial in $\Delta_\pm$ or equivalently in $\nu$. 

We now extract the large mass asymptotics in order to determine $\text{Pol}(\n)$. Using the $\Delta \to \infty$ asymptotics of the Hurwitz zeta function (see e.g.\ appendix A of \cite{HurwitzZeta})
\begin{equation}
 \zeta(s,\Delta) \simeq \frac{1}{\Gamma(s)} \sum_{k=0}^\infty \frac{(-1)^k B_k}{k!} \frac{\Gamma(k+s-1)}{\Delta^{k+s-1}} \, ,
\end{equation}
with $B_k$ the Bernoulli numbers, we find that for $\nu \to \infty$ (i.e.\ $m\to \infty$):
\begin{equation} \label{logZexp1}
 \log Z \, - \, {\rm Pol}(\nu) = \left( \log \nu^2 - 3 \right) \nu^2 - \frac{1}{12} \log \nu^2 \, + \, {\cal O}(1/\nu) \, .
\end{equation}
On the other hand the large $m$ heat kernel expansion (\ref{largemassexp}) gives in two dimensions
\begin{equation}
 \log Z = \frac{1}{4 \pi} \int_{S^2} d^2x \sqrt{g} \, \left( \int \frac{dt}{t^2} \, e^{-m^2 t}
 + \frac{R}{6} 
 \int \frac{dt}{t} \, e^{-m^2 t} \right) \, + \, {\cal O}(1/m) \, .
\end{equation}
After regularizing by introducing a lower cutoff $t > e^{-\gamma} \Lambda^{-2}$ with $\gamma$ the Euler constant, and dropping an $m$-independent quadratically divergent term, this can be written as
\begin{eqnarray}
 \log Z &=& \frac{1}{4 \pi} \int_{S^2} d^2x \sqrt{g} \, \biggl( m^2 \log \frac{m^2}{e\Lambda^2} - \frac{R}{6} \, \log \frac{m^2}{\Lambda^2} \biggr) \, + {\cal O}(1/m) \nonumber \\
 &=& \frac{1}{4 \pi} \int_{S^2} d^2x \sqrt{g} \, \biggl( \frac{\nu^2 + \frac{1}{4}}{L^2} \log \frac{\nu^2}{e(L\Lambda)^2} + \frac{1}{4L^2} - \frac{R}{6} \, \log \frac{\nu^2}{(L \Lambda)^2} \biggr) \, + {\cal O}(1/\nu) \nonumber \\
 &=&
 \left[ \,\mbox{same expression as previous line with $\Lambda$ replaced by $\L_\text{RG}$} \, \right] \nonumber \\
 && + \, \, \frac{1}{4 \pi} \int_{S^2} d^2x \sqrt{g} \, \biggl( \frac{\nu^2 + \frac{1}{4}}{L^2} \log \frac{\L_\text{RG}^2}{\Lambda^2} - \frac{R}{6} \, \log \frac{\L_\text{RG}^2}{\Lambda^2} \biggr) \, + {\cal O}(1/\nu) \,.
\end{eqnarray}
Here we introduced an arbitrary renormalization scale $\L_\text{RG}$. Using $R_{S^2}=2/L^2$ and ${\rm Vol}_{S^2}=4 \pi L^2$, the first line in the last expression can be evaluated as
\begin{equation}\label{hehohum}
 \left(\log \nu^2 - 1\right)\nu^2 - \frac{1}{12} \, \log \nu^2 + \bigl(\frac{1}{12} - \nu^2 \bigr) \log (L\L_\text{RG})^2 \, .
\end{equation}
The second line is a logarithmic renormalization of the couplings $\rho$ and $\phi$ of the classical action $S_\text{cl.} = \int d^2 x \, \sqrt{g} \, \left( \frac{\rho}{L^2} + \phi R \right)$. We can drop these terms provided we replace the couplings in the classical contribution to the full partition function of the theory by their logarithmically running counterparts $\rho(\L_\text{RG})$ and $\phi(\L_\text{RG})$. 
 
Comparison between (\ref{logZexp1}) and (\ref{hehohum}) thus yields:
\begin{equation} \label{PolnuS2}
 {\rm Pol}(\nu) = 2\left(1 -\log(L\L_\text{RG})\right) \, \nu^2 + \frac{1}{6} \, \log(L \L_\text{RG})  \, .
\end{equation}
Notice in particular that all $\log \nu$ terms have canceled, leaving us indeed with a pure polynomial, as needed for our analyticity arguments to be valid. This appears to be a nontrivial test of our formalism.

Summarizing, our final result for the exact, renormalized one loop partition function is\begin{eqnarray}
 \log Z_{S^2} &=& {\rm Pol}(\nu) \, + \, \sum_{\pm} 2 \, \zeta'(-1,\Delta_\pm) - (2 \Delta_\pm - 1) \, \zeta'(0,\Delta_\pm) \, ,
\end{eqnarray}
with $\Delta_\pm$ given by (\ref{bababam}) and $\text{Pol}(\n)$ by (\ref{PolnuS2}).
The determinant of the massive Laplacian on the 2-sphere was computed e.g.\ in \cite{Spreafico} (excluding the eigenvalue $m^2$ of the constant mode and suppressing scale dependence). The result is given there in terms of a certain infinite product. At the renormalization scale $\L_\text{RG}=1/L$, our result numerically equals theirs.

\subsection{dS$_{d+1}$} \label{sec:dSd}

The above computation for dS$_2$ can be straightforwardly generalized to dS$_{d+1}$ (i.e.\ $S^{d+1}$). Instead of (\ref{sumr}), we have more generally:
\begin{equation} \label{logZDR}
\log Z = {\rm Pol} - \sum_{\pm,r \geq 0} Q(r) \, \log \left( r + \Delta_{\pm} \right) \,,
\end{equation}
where the degeneracy $Q(r)$ is a polynomial in $r$, obtained from (\ref{Dell}) and some binomial manipulations as
\begin{equation}
 Q(r) = \frac{2r+d}{d} {r+d-1 \choose d-1} = D^{(d+1)}_r \, ,
\end{equation}
with $D^{(d+1)}_r$ as in (\ref{Dell}). For $d=1$, we have $Q(r)=2r+1$, for $d=2$ this is $Q(r)=(r+1)^2$, and so on. The conformal dimensions $\Delta_\pm$ are as usual related to the mass by
\begin{equation} \label{Deltapmdef}
 \Delta_\pm = \frac{d}{2} \pm i \nu \, , \qquad \nu \equiv \sqrt{m^2-(d/2)^2} \, .
\end{equation}
It is easy to see from the definition of the Hurwitz zeta function that for any polynomial $Q(r)$, we have $\sum_{r=0}^\infty \frac{Q(r)}{(r+x)^s} = Q(-x + \delta_s) \, \zeta(s,x)$, where $\delta_s$ is an operator acting on a function $f(s)$ as\footnote{For example, $\sum_{r=0}^\infty \frac{r^2}{(r+x)^s} = (-x+\delta_s)^2 \, \zeta(s,x) = x^2 \zeta(s,x) -2 x \zeta(s-1,x)+\zeta(s-2,x)$.} 
\begin{equation}
 \delta_s f(s) \equiv f(s-1) \, .
\end{equation} 
Using this observation, we zeta function regularize (\ref{logZDR}) as
\begin{equation} \label{logZdSd}
 \log Z_{S^{d+1}} = {\rm Pol}(\nu) + \sum_{\pm} Q(-\Delta_\pm + \delta_s) \, \zeta'(s,\Delta_\pm) |_{s=0} \, .
\end{equation}
This is just a polynomial in $\nu$ plus a finite number of standard Hurwitz zeta functions. 

To fix the polynomial ${\rm Pol}(\nu)$, we compare the large $\nu$ asymptotics of our expression (\ref{logZdSd}) to the asymptotics obtained from the heat kernel coefficients. For the sphere these are all computable, see for example appendix A of \cite{Birmingham:1987cy} for some explicit tables. By matching asymptotics, we find that (\ref{logZdSd}) is in fact the \emph{exact} result for odd $d+1$, that is to say, 
\begin{equation}
 {\rm Pol}_{S^{d+1}}(\nu) = 0 \qquad \mbox{(d+1 odd)} \, . 
\end{equation}
Interestingly, the large $\nu$ asymptotic expansion turns out to be purely polynomial in these cases, without any $1/\nu$ corrections (only with exponentially small corrections). The first few cases are
\begin{equation}
 \log Z_{S^3} \simeq \frac{\pi \nu^3}{3} \, , 
 \qquad \log Z_{S^5} \simeq - \frac{\pi}{12} \bigl( \frac{\nu^5}{5} + \frac{\nu^3}{3} \bigr) \, ,
 \qquad \log Z_{S^7} \simeq \frac{\pi}{90} \bigl( \frac{\nu^7}{4} + \frac{\nu^5}{3} + \frac{\nu^3}{270} \bigr) \, .
\end{equation}
Dividing these expressions by the sphere volume gives the (negative of the) renormalized one loop free energy density. Comparison to analogous computations of the free energy density in AdS (by our methods or as tabulated in appendix A of \cite{Hartman:2006dy}) shows that the results in the last equation are basically the naive analytic continuations of their AdS counterparts. 

For $d+1$ even, the situation is slightly more complicated. The polynomial part ${\rm Pol}(\nu)$ does not vanish and depends on the renormalization scale $\L_\text{RG}$. For example for $d+1=2$ we had (\ref{PolnuS2}) and for $d+1=4$, we get 
\begin{eqnarray}
 {\rm Pol}_{S^4}(\nu)=\bigl(-\frac{2}{9}+\frac{1}{6} \, \log(L\L_\text{RG}) \bigr) \, \nu^4 + \frac{1}{12}(-1+\log(L\L_\text{RG})) \, \nu^2 - \frac{17}{1440} \, \log (L\L_\text{RG}) \, .
\end{eqnarray}
Furthermore, the large $\nu$ asymptotic expansion of the full $\log Z$ does have an infinite series of $1/\nu$ corrections. 

The result (\ref{logZdSd}) appears to be a simpler expression than any we have been able to find in the literature for the determinant of a massive scalar on a $d+1$ dimensional sphere (although it is not hard to derive this formula from the conventional expressions for determinants). What made things simple here is the constraint of holomorphicity in $\nu$, and the fact that unlike the eigenvalues, the quasinormal frequencies are linear in the quantum numbers, leading to only elementary zeta functions.

\section{Numerical evaluation and WKB}
\label{sec:wkb}

For many spacetimes of physical interest, including the dyonic black holes we considered in \cite{us}, closed form analytic expressions for the quasinormal frequencies do not exist, and one has to resort to numerical computations or approximation schemes. Because our formulae expressing the one loop determinant as a product over quasinormal modes are all UV divergent, this poses an immediate practical problem: How can we extract the UV divergences and obtain the correct, finite, renormalized determinants? 

In principle, WKB type approximations can be used to control the UV tails of quasinormal mode series to arbitrarily high order in inverse powers of suitable quantum numbers. Examples of such computations include \cite{SW,Motl:2003cd,Natario:2004jd,Festuccia:2008zx,Dolan:2009nk}. Combined with numerical computations of the low lying modes which are outside of the WKB regime, this can be used in principle to compute the renormalized one loop determinant to arbitrarily high accuracy. 

A procedure along these lines was developed in \cite{DHLM} for the computation of determinants of flat space Laplacians with radially symmetric potentials. In their approach, the computation of the UV, high angular momentum ($\ell > L$) tail of the determinant is reduced to the evaluation of a number of integrals of local expressions similar to heat kernel coefficients, yielding an expansion in powers of $1/L$, in principle to any desired order. The UV divergences are extracted in analytic form and canceled by local counterterms in the usual way. The IR, low angular momentum part ($\ell < L$) is computed by applying the Gelfand-Yaglom formula \cite{GY,coleman} to each individual angular momentum mode, which reduces the computation of the determinant of the radial differential operator at fixed $\ell$ to solving a second order linear ODE. Combining the two parts produces a value for the renormalized determinant which becomes exact in the limit $L \to \infty$, and highly accurate already at modest values of $L$ provided the UV part is computed to sufficiently high order in $1/L$.

A similar procedure should be possible in principle for efficient and accurate numerical evaluation and renormalization of expressions such as (\ref{QNMfinalprodform}). Presumably the UV part can again be expressed in terms of local quantities, while the IR part must be computed numerically mode by mode. The IR part will often produce the qualitatively interesting physics; a number of examples are given in the discussion section.

\section{Generalization to other quantum numbers} \label{sec:otherquantumnumbers}
\label{sec:generalize}

If there are other conserved quantum numbers such as a momentum, angular momentum or Landau level, we can derive a generalization of equation (\ref{QNMfinalprodform}) where this quantum number takes over the role of the thermal frequencies. Let's call the quantum number $k$, and assume it can take values in some discrete index set, $k \in K$. Acting on eigenstates of the operator corresponding to this quantum number, the differential operator $D$ of which we wish to compute the determinant can be written as $D(k)$, where the part of the operator corresponding to the quantum number $k$ gets replaced by a c-number determined by k.\footnote{As a trivial example, if $D=-\partial_\tau^2 - \partial_x^2$ and $x$ takes values on a circle of radius $R$, the $x$-momentum $k$ takes values in $\frac{1}{R} \IZ$ and $D(k)=-\partial_\tau^2 + k^2$. \label{examplefn}} The operator $D(k)$ can be thought of as acting on scalars living on a dimensionally reduced space, with no explicit appearance anymore of the coordinate dual to $k$ (in the black hole case considered previously, this coordinate was $\tau$). However, the requirement of regularity on the original space can lead to particular additional boundary conditions on the reduced space. These could have some nonanalytic dependence on $k$ (in the black hole case we had $\phi \propto \rho^{|n|}$). Let us assume they are piecewise analytic on subsets $K_\alpha$ of $K$, $\alpha=1,2,\ldots,N$ (the analog of the positive and negative thermal frequencies). 
 
Now we drop the constraint $k \in K$ and look for pairs $(k_\star,\phi_{\alpha,k_\star})$, with $k_\star$ a to be determined complex number, solving the equation of motion
\begin{equation}
 D(k_\star) \, \phi_{\alpha,k_\star} = 0 \, , 
\end{equation} 
subject to the analytic continuation of the boundary conditions appropriate to the subset $K_\alpha$. This is the analog of the quasinormal mode equation (\ref{QNMEOM}). We denote the solution set of all $k_\star$ solving this problem by $K_{\star,\alpha}$.

For generic complex $k_\star$ not in $K$, $\phi_{k_\star}$ will not lead to a well defined function when lifted from the reduced to the original space, as it violates the quantization condition. Only when $k_\star \in K$ do we have a well-defined function on the original space, and in that case $\phi_{k_\star}$ is an actual zeromode of $D$. For typical real values of the parameters of the operator $D$ (such as the mass), there will be no such zeromodes. However after analytic continuation of these parameters to the complex plane, zeromodes $k_\star \in K$ can be expected to occur at complex codimension one loci in parameter space, which are zeros of the determinant. Denoting the degeneracy of the quantum number $k$ by $d_k$ and by matching poles and zeros as before, we are led to the generalization of (\ref{QNMfinalprodform}):
\begin{equation}
 \det D = e^{-\rm Pol} \prod_\alpha \prod_{k_\star \in K_{\star,\alpha}} \prod_{k \in K_\alpha} \, (k-k_\star)^{d_k} \, ,
\end{equation}
to be understood in a suitably regularized way, and with ${\rm Pol}$ a polynomial in the appropriate analytic parameters, determined by matching e.g.\ large mass asymptotics. 

As a trivial application, we can rederive the determinant for $S^n$ without knowledge of the quasinormal modes. We take $D=-\nabla^2 + m^2$ and pick as our quantum number the  angular momentum $\ell$. Then $D(\ell)=\ell(\ell+n-1) + m^2$ and $\ell_\star = -\Delta_\pm$ defined in (\ref{Deltapmdef}).  Furthermore $d_\ell=D^{(n)}_\ell$ is given explicitly in (\ref{Dell}). Thus
\begin{equation}
 \log \det D = -\rm Pol +  \sum_{\pm,\ell \geq 0} D^{(n)}_\ell \, \log(\ell+\Delta_\pm) \, .
\end{equation}
This is identical to (\ref{logZDR}), so the remainder of the computation is the same as in the remainder of section \ref{sec:dSd}. Written in this way, it is clear that this can also be formally obtained from $\log \det D = \sum \log \lambda$ by factorizing the eigenvalues $\lambda(l)$, with ${\rm Pol}$ representing a possible factorization anomaly.

\section{Discussion}

In this paper we have advocated an approach to computing determinants in thermal spacetimes
using sums over quasinormal modes, combined with analyticity in $\Delta$ arguments
to fix a `local' or `UV' contribution. We will now discuss possible physical applications of this approach
followed by a discussion of technical directions in which to build upon our results.

In applying the holographic correspondence to strongly
coupled field theories at finite temperature one may occasionally be interested in the contribution
of only a few quasinormal modes closest to the real axis. Such poles can dominate
the low energy spectral functions of the theory. Examples of situations where this occurs include
hydrodynamic poles \cite{Policastro:2002se, Son:2007vk}, cyclotron resonances \cite{Hartnoll:2007ih, Hartnoll:2007ip}, near the onset of bosonic instabilities such as superconductivity \cite{Gubser:2008px, Hartnoll:2008vx, Hartnoll:2008kx} and in the vicinity of Fermi surfaces \cite{Lee:2008xf, Liu:2009dm, Cubrovic:2009ye, Faulkner:2009wj}. It was the last of these that initially motivated us to develop the
formalism presented in this paper \cite{us}, but we expect all of the examples just listed to lead
to nontrivial physics at the one loop level.

From the perspective of the holographic correspondence, we
have expressed the $1/N$ correction to the free energy of a strongly coupled field
theory in terms of the poles of retarded Green's functions of operators in that theory. This may
suggest a reformulation of the finite temperature theory with the poles of Green's functions as
the `elementary' quantities.

We obtained various exact expressions for scalar determinants. Of these the thermal $AdS_3$ and BTZ
results matched previously obtained expressions while the compact formula (\ref{logZdSd}) for de Sitter space in arbitrary dimensions, in terms of a finite sum of elementary Hurwitz zeta functions, appears new (to us). Similar results can be obtained for thermal AdS in arbitrary dimensions. Furthermore, determinants in topological black hole backgrounds should be computable in essentially the same way as the BTZ case; the relevant quasinormal modes have been found in \cite{Birmingham:2006zx}. A recent list of other spacetimes for which the quasinormal modes are known analytically can be found in the introduction of \cite{LopezOrtega:2009zx}.

We briefly outlined a scheme for efficient numerical evaluation of renormalized determinants as products over quasinormal modes in cases where analytic expressions do not exist, by combining UV WKB results with IR numerics. We leave implementation of this to future work.

Important generalisations of the formalism include the extension to stationary, rather than static, spacetimes and to more general asymptotic geometries. This should proceed along the general lines sketched for arbitrary quantum numbers in section \ref{sec:otherquantumnumbers}, but some additional care may be needed e.g.\ in doing the Wick rotation or making analyticity arguments. Indeed, at a more formal level it would be of interest to prove (or disprove) that the various determinants we are computing indeed have the strong analyticity properties we have assumed.

\section*{Acknowledgements}

All the authors are pleased to acknowledge the hospitality of the KITP in Santa Barbara while this work was being completed.  This research was supported by the National Science Foundation under grant DMR-0757145 (SS), by the FQXi foundation (SAH and SS), by a MURI grant from AFOSR (SS), and by DOE grant DE-FG02-91ER40654 (FD and SAH). The research at the KITP was supported in part by the National Science Foundation under Grant No. PHY05-51164.

\end{document}